# Bibliometric Approximation of a Scientific Specialty by Combining Key Sources, Title Words, Authors and References

**Nadine Rons** [a]

[a] Vrije Universiteit Brussel (VUB), Research & Data Management Dept., Pleinlaan 2, B-1050 Brussels, Belgium

*E-mail address:* Nadine.Rons@vub.ac.be

ABSTRACT

Bibliometric methods for the analysis of highly specialized subjects are increasingly investigated and debated. Information and assessments well-focused at the specialty level can help make important decisions in research and innovation policy. This paper presents a novel method to approximate the specialty to which a given publication record belongs. The method partially combines sets of key values for four publication data fields: source, title, authors and references. The approach is founded in concepts defining research disciplines and scholarly communication, and in empirically observed regularities in publication data. The resulting specialty approximation consists of publications associated to the investigated publication record via key values for at least three of the four data fields. This paper describes the method and illustrates it with an application to publication records of individual scientists. The illustration also successfully tests the focus of the specialty approximation in terms of its ability to connect and help identify peers. Potential tracks for further investigation include analyses involving other kinds of specialized publication records, studies for a broader range of specialties, and exploration of the potential for diverse applications in research and research policy context.

*Keywords:*

Scientific specialty

Scholarly communication

Bibliometrics

Scientific peers

*Funding:*

This research did not receive any specific grant from funding agencies in the public, commercial, or not-for-profit sectors.

# 1. Introduction

*1.1. Research policy context*

Science and innovation policies increasingly focus on highly specialized subjects such as individual scientists, specific research programmes, research proposals, emerging specialties, and scientific breakthroughs. The related 'specialties' are coherent sets of subject-related research problems and concepts, focused on by an interacting research community of specific scientists and research teams. To answer many kinds of concrete questions posed at this level, research management and evaluation procedures typically seek advice from peers belonging to the particular specialty, possibly supported by quantitative material. Both for the identification of peers and for the production of quantitative material, bibliometric information and techniques can be helpful if these can sufficiently adequately capture the particular specialty. The aim of the method proposed in this paper is to produce a set of publications that approximates a subject's specialty sufficiently closely for such practical purposes in research management (not to generate an exhaustive representation of a specialty or to define it). This paper presents the general principles of the method, already indicating certain possibilities for further development, tuning and potential applications. Further research targeting particular practical purposes can subsequently lead to translations into protocols or tools for research policy makers and administrators.

*1.2. Bibliometric context and debate*

Sufficient bibliometric focus at the specialty level requires an aggregation of publications that is more fine-grained than the broad subject categories grouping interrelated journals (Pudovkin and Garfield, 2002) that form the backbone of the commonly used global publication and citation indexes (Clarivate Analytics' - formerly the IP & Science business of Thomson Reuters - Web of Science, and Elsevier's Scopus). A need to study the fine structure of science was felt already soon after the development of the first science citation index (Garfield, 1955), as summarized by Small and Griffith (1974). Recent studies still point to observations of important internal heterogeneity within broad disciplines, in general (Zitt, Ramanana-Rahary, and Bassecoulard, 2005) and within a wide-spread variety of specific domains (Chemistry: Neuhaus and Daniel, 2009; Engineering: Lillquist and Green, 2010; Economics: van Leeuwen and Calero-Medina, 2012; Medical subject categories: Schmidt and Sirtes, 2015; Library and Information Science and Science & Technology Studies: Leydesdorff and Bornmann, 2016). These observations call for caution in interpreting indicators and propose to use more fine-grained classifications at the specialty level. Specialties described in various terms have been the subject of studies from sociological, bibliographical, communicative, and cognitive approaches (as for example reviewed in a specialty mapping context by Morris and Van der Veer Martens, 2008), but a standard, generally applicable fine-grained method that can sufficiently closely approximate the specialty of a single scientist or a team to represent its specific publication and citation characteristics, has not yet been established. In its absence, broad subject category structures are still used in analyses concerning much more specialized entities. In the particular context of individual scientists, recent debates on challenges and ethical issues in bibliometric applications mainly focus on properties following from the design of current indicators. An adequate delineation of specialties, drawing a frame within which investigated entities and their performances can be confidently positioned, is a less debated but equally important issue (Leydesdorff and Bornmann, 2016).

*1.3. Proposed method and paper outline*

This paper presents a novel method that approximates the scientific specialty to which a given highly specialized 'seed record' of publications belongs. Four sets of key values are determined from the seed record (for the data fields containing source, title, authors and references) and partially combined to approximate the specialty by the set of publications associated to key values determined for at least three of the four data fields. The selection of the four data fields is founded in concepts defining disciplines, and the determination of key values is founded in empirically observed regularities.

A preliminary test of the method's added value compared to coarser domain delineations was the verification of its capability to keep related specialties apart. This test was successfully performed for two scientists in Theoretical and Experimental Particle Physics, closely related specialties in terms of publication venues but strongly differing in other publication and citation characteristics (Rons, 2016). The case used in the present paper to concretely illustrate how the method works adds a second, complementary test, verifying whether produced results can be confirmed to belong to the specialty. This paper's test material consists of known peers in a specialty in the domain of Biology. Both tested abilities are essential for the method's aim. An extension to other types of cases and other specialties requires further research and possibly different operationalizations than the ones used in this paper.

The next section describes the method's conceptual and empirical foundations, situated in diverse areas in philosophy of science, information science and linguistics. It refers to particular literature from these areas that addresses the aspects discussed and built upon. It is followed by the section describing the different phases of the method itself, discussing the choices that were made from different options for the method in general and for the particular application shown in this paper. The principles of the method and the choices described are not complex, enabling the interested researcher in the field to further explore the method's potential and limitations for different purposes (the overall complexity of an investigation will also depend on specificities of the chosen publication database). The section on the illustrated application and data describes how the application is used to test the method, its relevance in a research policy context, and the sample of peers used. The results section describes the results obtained in the different phases of the method and the implications for the performed test. The discussion section summarizes the outcome of the test of the method, and points to a range of potential application areas and to possibilities for more advanced criteria designs than those used in this paper in a 'proof of concept' context.

**2. Foundations**

*2.1. Conceptual foundations*

The term 'specialty' has been used in literature for different levels of aggregations of publications. In this paper it refers to a specialty as regarded by Braam, Moed, and van Raan (1991, p. 234, at the introduction of a combined co-citation and word analysis approach) from a problem-solving perspective (Laudan, 1977): a "coherent set of subject-related research problems and concepts upon which attention is focused by a number of scientific researchers", irrespective of their social and intellectual backgrounds. The associated "self-organized network of researchers" tends "to study the same research topics, attend the same conferences, read and cite each other's research papers and publish in the same research journals" (Morris and Van der Veer Martens, 2008, p. 214-215). These definitions correspond to a scientific specialty as typically dealt with in research policy governing a specific community of researchers (e.g., in a university): a body of research on closely related scientific problems, building on a largely common body of resources (previous work including theories, methods, models, ...), and mainly produced by scientists who can be regarded as each other's peers (having knowledge of the appropriate and most esteemed publication venues in the area) and potential reviewers (to an extent depending on the proximity of their personal research lines and broader experience gained in their careers). The body of research and the body of resources built on may be largely the same (for example for long established 'closed' specialties) or may differ (for example for interdisciplinary research and emerging specialties). Specialties may evolve, emerge and disappear over time. Specialties may also split up when increased complexity necessitates a research community to focus on several particular parts of problems, in its research, its communication channels, and its practical organization. This dynamic nature of science and its research community structures, repeatedly reorganizing itself and its contributions to an ever increasing volume of scientific knowledge, can be made apparent in bibliometric visualizations of timelines (Chen, Ibekwe-SanJuan and Hou, 2010), of threads and isolates constituting a specialty (Klavans and Boyack, 2011) and of term usage in a domain over time (Milojevic et al., 2011). This dynamic nature may also partly explain why the concepts of specialties, fields, disciplines, domains, ... are often ill-defined in studies, the terms being used sometimes interchangeably and sometimes in hierarchy.

From an extensive review of literature, Sugimoto and Weingart (2015) synthesized a framework of six criteria defining disciplines. The same notions and conceptualizations also tie together more strongly focused specialties as defined above. To proceed towards a basis for bibliometric criteria to approximate a specialty, publication data fields were matched to these conceptual components, and to the four facets of the framework for bibliometric analysis of scholarly communication as proposed by Ni, Sugimoto, and Cronin (2013) (**Table 1**). The latter framework extends Borgman's (1989) three-faceted framework with gatekeepers. In this conceptual context it was also stressed that measures of the scholarly landscape based on a single facet produce partial results and that a combination of multiple facets is needed for full understanding.

**Table 1**

Publication data fields matched to conceptual components defining disciplines and scholarly communication.

| **Notions and conceptualizations defining disciplines** (Sugimoto and Weingart, 2015) | | **Matching publication data fields** (principal match) | **Facets of scholarly communication** (Ni, Sugimoto, and Cronin, 2013) | |
|---|---|---|---|---|
| Separatedness | Homogeneous domain with specific perspectives, methods, theory, knowledge, distinguishing from and not generally shared with other knowledge formations | Source[a] (journal or other) | Gatekeepers | Selection for publication reflecting the publication channel's scope and standards |
| Communicative | Specific terminologies, specific technical language | Title[a] Keywords | Concepts | Word choices and assigned indexing terms |
| Social | Organized community of researchers who associated to facilitate intercommunication and establish standards of inquiry and authority thereof | Authors[a] | Producers | Individual authors or aggregations of authors |
| Cognitive | Common body of intellectual content, theories, methods, models, ..., coherently organized | References[a] | Artifacts | Publications or aggregations of publications |
| Tradition | Dynamic nature of knowledge communities, stages of development, matured (showing continuity, persistence and stability over time) versus emerging | Publication year | — | |
| Institutional | Authority, formality, identity, stability (perpetuating by training, supervising conduct, validating competence, ensuring responsible use of competence, degree-granting, journal development) | Addresses | — | |

[a]: *Publication data fields selected as a basis for bibliometric criteria to approximate specialties*

A set of four publication data fields was selected as a basis on which to define bibliometric criteria to approximate a specialty: source, title, authors, references. Values in these four selected data fields are generally available for scholarly publications in the global publication and citation databases, and cover all four components of the framework defining scholarly communication (gatekeepers, concepts, producers, artefacts) and four of the six components of the framework defining disciplines (separatedness, communicative, social, cognitive). The concepts covered in each of the two frameworks are mainly "inward looking" concepts related to intrinsic properties (specific common body of knowledge, research community, terminology), but also include "outward looking" concepts related to boundaries that keep literatures separated and free of inappropriate contributions (separatedness, gatekeepers).

The two components of the discipline-defining framework that are not covered by the four selected publication data fields (tradition, institutional) are matched to data fields related to positions in time and space (publication year, address). An intrinsic difference these have in common with respect to the four selected publication data fields is that publication year and address are not related to active choices by an author preparing a particular publication, but rather follow from evolutions, progress made, and opportunities encountered. While particular values of these data fields can limit the possibly associated specialties, the values as such do not indicate the specialty of a publication (publication year) or are in general less specifically associated to the publication's specialty than values of the selected data fields (address). Concerning usage of the address field to indicate a publication's research area, counter-indications have been reported for applications at more aggregated levels than research specialties. Bourke and Butler (1998) found that, while department names in the address field designate the close environment in which scientists produce publications, these may only poorly correspond to the names of research fields to which the publications contribute. In this paper's more focused context aiming to distinguish between specialties, department level information likely is even less suitable, because of the generally wider scope of a department and the non-systematic availability at present in publication databases as well as on publications indexed. A department name often is not included in an author's affiliation and typically points to a wider discipline in which the publication's specialty is situated, including closely related specialties that it needs to be distinguished from. Neither the publication year, nor the address field in its present state, were therefore included in the set of data fields to help approximate a specialty. The publication year (and its potential time-related influence on the content of the specialty approximation) is present in the method in a different way, via the time frames chosen for the seed record and for the constructed specialty approximation.

Two publication data fields match a same component in each of the two conceptual frameworks: keywords and title. From these two only the title was included in the set of selected publication data fields. Title words have been used before to frame a publication's scientific content, for example in general automatic indexing systems (Neufeld, Grahams, and Mazella, 1974), and for the delineation of research areas ranging from social sciences (Nederhof and Van Wijk, 1997) to biomedical sciences (Lewison, 1999). A title provides the most specific indications of a document's content in publication lists. It therefore offers a crucial opportunity for authors to point out the particular contribution of a publication, and to make it eye-catching and interesting enough for the targeted audience to decide to read further. Also editors recommend that titles should convey not only the topic, but also more specific aspects such as study design, methods, results and conclusions (Goodman, Thacker, and Siegel, 2001). As a result, a high lexical density is observed in titles of all scientific publication types (Gesuato, 2008). Journal article titles in particular have become more informative and specific over time (Berkenkotter and Huckin, 1995), containing more content words for example identifying new techniques and aspects studied (Buxton and Meadows, 1977) and more complex words reflecting increasing field complexity (White and Hernandez, 1991). This makes title words a highly suitable basis to associate publications to a specialty, in particular in recent periods. Also keywords may include highly specific terminology, but are often also broader labels under which the research can be classified (one of the specialty's methods or theories, or a particular kind of subject these are applied to). Whittaker, Courtial, and Law (1989) pointed out this conceptual and functional difference between title words (emphasizing a publication's originality) and keywords (relating it to other publications). From a comparative analysis they conclude that title words are rather suitable for the examination of small document sets or closely defined areas of science (this paper's context) and keywords are to be preferred for larger or more heterogeneous data sets. Title words further have the advantage over keywords (and abstracts and full text) to be more generally available for scholarly publications in global publication databases such as Web of Science and Scopus. Using title words rather than these alternatives thus optimizes the share of publications that can fully contribute to an analysis, and avoids source related bias that could result from differences in publishers' policies regarding data made available.

Parallels with the combination of the four data fields selected for the proposed specialty approximation method can be observed in the way scientists look for literature that is of interest to them for their current and future work. They also need to combine different aspects to obtain adequate results, typically spanning the same four dimensions as those combined in this paper, based on sources, words, authors and references/citations. Important criteria used in deciding what to read from the available literature and related document information elements providing most clues (Wang and Soergel, 1998; Tenopir et al., 2011) are a publication's topical relevance, scientific orientation/level and novelty (informed primarily by its title) and a publication's quality (informed primarily by its source and authors). References and citations provide links to subsequent potentially interesting publications. Starting for instance from a newly found publication of interest, a scientist may search for other publications of interest (1) by scanning the publications referred to by the publication, looking for familiar title words, authors, or sources that are known to indicate potentially interesting contributions to the specialty, and (2) by looking up publications referring to those that were already found to be of interest. Mimicking such combinations made in human search operations has for example inspired reading recommendation strategies. Steinert, Chounta, and Hoppe (2015) combined textual and citation network based similarities, a combination that had been advocated already by Pao (1993) in a context of information retrieval, and that was found to outperform recommendations to cite based on textual similarity measures only (Strohman, Croft, and Jensen, 2007). Also the specialty approximation method proposed in this paper could be seen as a mimicking of human search operations, on a wider four-dimensional scale.

*2.2. Empirical foundations*

For each of the four selected publication data fields, using criteria for that data field only can correctly identify a large part of a specialty's publications, but results may substantially suffer from false positives and false negatives. For the source field this is evident from Bradford's law, observing that articles on a given subject are published in a nucleus of periodicals more particularly devoted to the subject, and with smaller productivities in several other groups of journals (Bradford, 1934). Analogous phenomena can be expected for references, authors and title words: research results from a particular specialty to a large extent, but not exclusively, refer to a common body of intellectual content for that specialty, are published by an organized community of authors in that specialty, and use a specific terminology for that specialty. This non-exclusiveness implies an inherent interconnectedness between specialties and makes it an extremely challenging endeavour to design a systematic way to delineate research specialties.

To limit false positives and false negatives is found highly difficult in practice when focusing on only one dimension and its corresponding basic technique(s) such as word-frequencies (Luhn, 1958), co-word analysis (Callon et al., 1983), bibliographic coupling (Kessler, 1963), and co-citation (Marshakova-Shaikevich, 1973; Small, 1973b). The difficulties thus encountered have been discussed since long (Rip, 1988; Leydesdorff, 1997). The abilities of techniques to interconnect individual publications and to associate publications to a particular research area have been amply compared and combined, sequentially as well as integrated in various ways. Examples can be found in a wide range of contexts, including science mapping or clustering of scientific fields (Braam, Moed, and van Raan, 1991; van den Besselaar and Heimeriks, 2006; Janssens, Glänzel, and De Moor, 2008; Zitt, Lelu, and Bassecoulard, 2011), information retrieval (Pao, 1993), delineation of particular scientific fields (Zitt and Bassecoulard, 2006), topic discovery (Chikhi, Rothenburger, and Aussenac-Gilles, 2008), research front representation (Boyack and Klavans, 2010), identification of core documents used to represent document clusters and topics (Glänzel and Thijs, 2011) and to detect emerging topics (Glänzel and Thijs, 2012), reference sets for normalization (Colliander, 2015), retrieval of highly related articles (Liu, 2015), computation of similarity of papers (Hamedani, Kim, and Kim, 2016).

A general tendency in conclusions is that hybrid methods that combine different dimensions generate superior results. It also appears that methods that report high accuracies in capturing contributions to a particular specialty would not be easy to apply on a daily basis to ad hoc cases in research management because of a relatively high complexity, computation or labour intensiveness, or required participation by experts in the investigated domain. The method proposed in this paper seeks to generate sufficiently adequate results via a series of steps of relatively low complexity. Starting from a seed record, it makes use of regularities observed in the occurrences of sources, title words, authors and references to determine key values for these fields, and subsequently build a specialty approximation using these key values. As in a wide variety of natural and man-made phenomena, a relatively large part of produced results / effects (here publications) stem from a relatively small part of occurring contributors / causes (here values in the selected data fields). This is known as the Pareto principle, suggested in 1937 by Joseph Juran to separate the "vital few" resources from the "useful many". Such observations can often (but not always) be adequately modelled by power laws (Clauset, Shalizi and Newman, 2009). In the case of publications, the regularities observed in the four selected data fields are mostly described by well-known empirical laws revealed early in the 20th century, recently discussed in historical context by De Bellis (2014) and Furner (2016).

(1) Sources: It can be expected from Bradford's law (Bradford, 1934), as reformulated by Garfield (Garfield, 1971), that a limited number of sources contain a substantial share of publications of a highly specialized publication record.
(2) Title words: It is clear from Zipf's law (Zipf, 1935) that certain title words are more suitable than others to frame the content of a specialty.
(3) Authors: Lotka (1926) examined frequency distributions of scientific productivity of senior authors in two disciplinary databases, and concluded that "the number making $n$ contributions is about $1/n^2$ of those making one; and of all contributors, the proportion that make a single contribution, is about 60 percent". Lotka's law concerns not total individual productivity but individual productivity in particular fields (MacRoberts and MacRoberts, 1982). As the scope of the proposed specialty approximation is similar to the focused productivity context of Lotka's findings, a limited percentage of senior authors can be expected to produce a large percentage of publications in a specialty. In this perspective, key authors may be determined among reprint authors when this information is sufficiently systematically available and representative for leading authors (even when not necessarily senior authors). In particular in areas with high typical numbers of authors per publication, this approach can limit key authors to a smaller, more efficient selection.

(4) References: A regularity with comparable effects, although not formulated as a law, can be expected from long-standing knowledge concerning citations and from more recent observations of references. Distributions of citations received by publications published in a particular year are well known to be highly skewed (Narin and Hamilton, 1996). Consequently, this unevenness is also present in the distribution of references given in publications published in a particular year over documents referred to. Such frequencies of occurrence of references cited by publications in particular research specialties have been examined to detect seminal literature (Kostoff and Schlesinger, 2005; Kostoff et al., 2006; using Citation-Assisted Background) and historical roots (Marx et al., 2014; using Reference Publication Year Spectroscopy). To proceed towards similar observations of frequency distributions as those underpinning the selection of key values for the other fields, the same data can be observed from a different perspective. The studies by Kostoff and Schlesinger (2005), Kostoff et al. (2006) and Marx et al. (2014) identify publications that are frequently referred to as compared to contemporaries in nearby publication years, to indicate the specialty's seminal literature in a historical perspective. Another perspective (the one in this paper) is to detect references most representative for a specialty's research that is published in a particular studied time frame, requiring the identification of publications that are frequently referred to as compared to other publications referred to in that studied time frame. This perspective was already investigated by Price (1965), drawing a picture where references from publications in a single year for one half refer to about half of all publications published in previous years, and for the other half refer to a quite small group of earlier publications (based on the observed distribution of numbers of references in publications placed in a context of a growing world literature). The regularities in frequencies of occurrence of individual references then suspected by Price, can also be expected from numbers of references to publications most frequently referred to as published by Kostoff and Schlesinger (2005), Kostoff et al. (2006) and Marx et al. (2014; looking at citation levels attained in the most recent year). While a more thorough study of these underlying regularities is out of scope of this paper, the test sample studied provides an illustration of how a relatively small fraction of references is referred to by a relatively large fraction of publications in the seed record (illustrated in section 5.2. Results in Phase 2: Key values), similar to findings for the key values for the three other data fields.

**3. Method**

Building on the described conceptual and empirical foundations, the proposed method consists of three consecutive phases: (1) specification of the highly specialized seed record that is the starting point of the analysis, (2) determination for each of the four data fields of a set of most frequently occurring values ('key values') characterizing the seed record, and (3) identification of all publications associated to key values for at least three of the four data fields, constituting the specialty approximation. These three phases are described in more detail below and visualized schematically in **Figure 1**. Depending on the type of analysis, all information needed may already be present in the key values determined in Phase 2 (e.g., to verify the degree of coverage of a given publication record by the specialty approximation), or the specialty approximation may effectively need to be constructed in Phase 3 (e.g., to identify strong contributors).

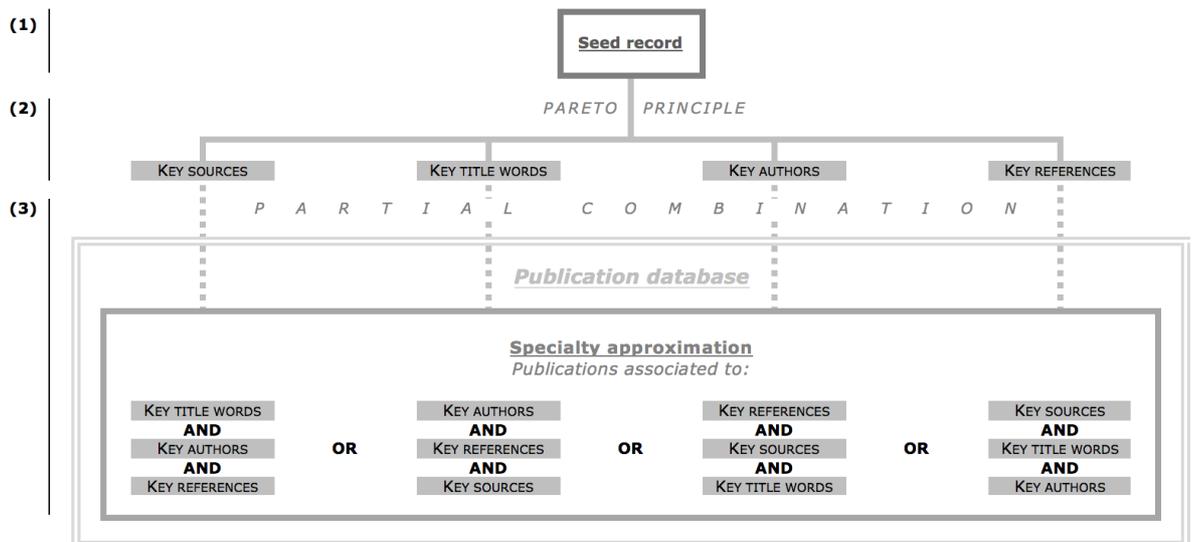

**Fig. 1.** Schematic visualization of the proposed method for bibliometric approximation of a research specialty by partial combination of key values for four publication data fields.

In each phase there is a possibility to restrict the intermediate or final results to certain document types. Among the document types indexed in the global publication and citation databases, a distinction is commonly made between those types that report original research results or summarize key research (article, letter, review, proceedings paper, and formerly also note) and other types that for instance rather represent interpretations or introductions (including meeting abstracts, editorial material, book reviews). The latter are avoided in assessments of scholarly contributions to the pool of knowledge, where inclusion of more diverse types of scientific communications would distort results. For the application shown in this paper, it was not expected that the presence of any particular document type would compromise the calculations or results. No document type was therefore omitted, also retaining maximal volume for the small scale seed record and more broadly accommodating the diversity of publication profiles among individual researchers.

*3.1. Phase 1: Specification of the seed record*

The seed record should be a well-understood set of publications that is trusted to (mainly) belong to the particular specialty for which an approximation is to be created. It may for instance originate from the publication record of an individual scientist, of a research team, or of a research programme. The considered publication period depends on the analysis' scope, for example spanning an entire research career or focusing on a limited recent period.

For the method to be successful, a seed record should sufficiently feature important references, authors, title words, and sources for the specialty to be approximated. Whether this can be expected for a particular publication record submitted for analysis, considering how it was obtained or constructed, is an important initial point of attention in view of adequate outcomes. Publication records of individual scientists for instance are inherently biased in one of the four data fields: authors. Such a publication record can however easily be enlarged to a seed record with a broader author representation but remaining focused on the specialty concerned: by adding the publications referred to. This procedure adds to each publication in the initial coherent publication record a coherent set of publications that is judged of interest to it in a coordinated action by its authors. The entire set of publications referred to can therefore be expected to mainly belong to the specialty of the initial publication record (at least in case of non-interdisciplinary research), and can be confidently used as an extension mainly associated to the same specialty. This intervention can be applied to any initial publication record where it is judged useful to increase volume and diversify values in data fields, enhancing its capturing of important sources, title words, authors and references for the specialty.

A similar enrichment of papers with information from citing as well as cited publications has been used in for example topics discovery (Chikhi, Rothenburger, and Aussenac-Gilles, 2008) and paper recommendation (Sugiyama and Kan, 2010). It is less evident however to maintain a same focus after enrichment from or extension with publications that cite an investigated publication record than it is with publications that it refers to. Citing publications to an important extent result from uncoordinated actions by unrestricted authors in possibly different disciplines, in unanticipated proportions. This reasoning was confirmed by observations of citing sentences, revealing that different citations to a same publication often focus on different aspects of that publication, including aspects not focused on in the publication's abstract (Elkiss et al., 2008). With respect to a particular publication or coherent group of publications, references and citations indeed embody different 'signs' (Cronin, 2000), respectively connecting to research situated in the past through acknowledgments given and to research situated in the future through acknowledgments received. Particularly relevant for the proposed method is the characterization of references and citations as respectively contextualized and decontextualized actions from the perspective of a particular publication (Wouters, 1998). From this observation it follows that, at the aggregated levels of all references versus all citations, these link to groups of publications of a possibly different composition as regards scientific focus in relation to the publication(s) started from. The degree of difference could in particular depend on the importance of the specialty of the publications started from in the knowledge base of other, relatively large specialties. This difference at the aggregated level makes references and citations intrinsically more suitable for different kinds of applications: references for grouping documents according to scientific focus, and citations for estimating a possibly diverse scientific impact. A choice between the two can easily be made when there is a particular set of publications as a starting point. This is the case for the method proposed in this paper: an enhanced seed record is constructed starting from a smaller publication record by adding the publications it refers to in Phase 1, and key references are determined starting from the seed record in Phase 2.

*3.2. Phase 2: Determination of key values characterizing the seed record*

The key values are selected starting with the value(s) corresponding to the highest frequency of occurrence and consecutively adding values corresponding to the largest remaining frequency of occurrence, until at least a predefined majority ('coverage threshold') of publications in the seed record is 'covered' by key values: associated to at least one key value determined for the references, authors and source fields, and to at least two key values determined for the title field, as explained in the respective subsections below. The coverage threshold sets a boundary between values less and more strongly associated to the specialty (by frequency of occurrence in the seed record), which in Phase 3 will require respectively more and less confirmation via association to key values for other data fields for a publication to be included in the specialty approximation.

The coverage threshold can be adapted to suit a particular application or investigated sample, and can be different for each of the four data fields. Higher thresholds can increase coverage of the specialty and lower thresholds can more strongly focus on its most frequently occurring elements. Either may be preferred depending on the aspect of primary importance in a particular application. If the coverage threshold can only be reached by descending to values occurring only once, then the appropriateness of the chosen coverage threshold or the adequacy of the seed record may need to be reconsidered, or special characteristics of the specialty may require adaptations to the method (to be investigated for instance for interdisciplinary specialties).

Because coverage by key values for one data field will be complemented in the method by coverage for at least two other data fields, less elaborate checks and conditions are required for each data field separately. In this paper's 'proof of concept' context, key values are determined in relatively uncomplicated ways as described in the paragraphs per data field below. These simple operationalizations have the advantage to be more easily replicated and tested by the interested researcher in the field than more advanced designs that could be incorporated in the method (of which some are indicated in the discussion section).

*3.2.1. Key cells of sources*

For the source field, not key sources but coherent key groups of sources are determined. This approach (1) avoids source related bias stemming from an individual scientist's publication and citation behaviour, (2) includes also other document types than journal publications (which may be important or even dominant in certain specialties, such as contributions in proceedings in Engineering and Informatics), and (3) facilitates inclusion of contributions in new journals (recently created or included in the database).

The groups of sources used in this paper are the partition cells defined by Rons (2012): each cell containing all sources associated to a same combination of subject categories (or, where a source's subject category association changed, the respective part of its publications). The term "cell" stems from mathematical terminology, where blocks, parts or cells of a partition of a set $X$ are nonempty subsets of $X$ such that every element $x$ in $X$ is in exactly one of these subsets. It was confirmed empirically for scientists in the domain of Mathematics and in subdomains of Physics (Rons, 2014; Rons, 2016) that a limited number of cells contain a substantial share of publications of a highly specialized publication record, similar to the findings for journals which constitute part of the sources in a cell. The key cells are determined from the distribution of seed record publications over the partition cells.

*3.2.2. Key title words*

For the title field, key values are determined among title words as the potentially reoccurring values among the publications in a specialty (while the title as a whole is primordially linked to exactly one publication). Only words that are associated to one specialty in particular and much less to other specialties are suitable to help distinguish between publications from different specialties. Non-suitable words, in particular 'stop words', are to be avoided. Such words with no important significance can differ depending on the context. There is therefore no single universal list of stop words for all applications.

As a simple operationalization in this paper, the method based on word length that was used previously in a context of subfield delineation (Lewison, 1999) was chosen and found to work sufficiently well in the studied cases (Rons, 2016; this paper) with a slightly more lenient threshold title word length: excluding words of less than five instead of less than six characters. Among the remaining words, key title words are determined from the frequency of occurrence of title words among seed record publications.

The presence in titles of words associated to the specialty, the degree of specificity of title words and word usage over time can however strongly vary. A particular challenge in this context of specialty approximation is to avoid that selected key words that are also frequently used in related specialties (possibly sharing certain journals and knowledge referred to) lead to the inclusion of a substantial amount of publications from those related specialties. Of additional importance in this context, is that a word that is unimportant standing by itself, even one included in a stop word list, might be part of an important word combination for the research specialty investigated (e.g., "second" in combination with "order" in mathematics). Also Lewison (1999) observed that titles can contain words that by themselves link to large parts of other specialties, and only effectively link to the specialty aimed for in combinations with other words. These observations lead to the requirement that at least two key title words should be present for a publication to be considered as 'covered' by key title words.

A similar choice to use two words (in word pairs) out of a set of terms from titles and abstracts was made in the analysis accompanying the introduction of the technique combining co-citation and word analysis (Braam, Moed, and van Raan, 1991). A higher required number of key title words may improve results for some specialties, but not for certain other specialties such as those with a relatively low number of words per title (Lewison and Hartley, 2005).

The described relatively simple way to restrict and select key title words could be substituted with more advanced techniques, as pointed to in the discussion section. Not all possible refinements are advisable however in this context of specialty approximation. A choice not to further reduce potential key title words to specialist terms or to word stems can be justified by the observation that disciplines — besides in specialist terminology — differ in socio-epistemic discourse, and can be studied by discourse epistemetrics in for example dissertation abstracts (Demarest and Sugimoto, 2015). Therefore also in publication titles, particular specialist and non-specialist word variants may be more strongly associated to one specialty than to other specialties, possibly as a component of a composite specialist expression. Maintaining a larger variety of types and variants of words may therefore better serve processes aiming to approximate a specialty, if combined with a required presence of several words from a selected list, such as the required two key title words in the method proposed in this paper.

*3.2.3. Key authors*

Key authors are determined from the frequencies of occurrence of authors in seed record publications, either among all authors or among a subcategory thereof such as reprint authors. The subcategory of reprint authors was used for the Particle Physics examples examined by Rons (2016) and also in this paper for a sample situated in the Biology discipline. A restriction to reprint authors may however not be equally suitable for all specialties (for example possibly less so for more interdisciplinary research).

'Unique' authors were identified by a name and first initial, and homonyms were avoided among key authors by excluding frequently occurring names (here roughly defined as names occurring in the seed record with more than one first initial, as in Rons, 2016). More advanced author disambiguation techniques are available and may be substituted as referred to in the discussion section. The chosen simple operationalization can be sufficient in this phase of key author determination from a specialized seed record because homonyms are less likely to occur than in less specialized areas, and an occasional merger of homonyms in the set of key authors would merely add extra authors who contribute to the specialty's knowledge base. The issue of homonyms however gains importance in the phase of actual construction of a specialty approximation. Even when the inclusion of publications by homonyms from outside the specialty is limited by the required coverage by key values for at least three data fields, extremely frequently occurring names can be a heavy burden in calculations.

*3.2.4. Key references*

Key references are determined from the frequencies of occurrence of references in seed record publications. No restrictions need to be set on the references' ages or document types, allowing maximal inclusion of the knowledge base used (accommodating also the slower publication and citation traditions of certain specialties).

To facilitate calculation certain restrictions can be imposed on the nature of references taken into account, provided that these do not result in a significant information loss (which may depend on the specialty). For instance, for specialties where a sufficiently important share of references is identified by DOI and indexed in the database used, key references can be limited to this subset. This was done for the Particle Physics samples used in Rons (2016) and also for the Biology sample used in this paper.

*3.3. Phase 3: Construction of the specialty approximation*

The four sets of key values determined in Phase 2, representing relatively small shares of occurring values for each data field, are combined to jointly more strongly characterize the seed record, and form a basis for approximating the specialty. Criteria based on each publication data field separately can generate 'false positives' and 'false negatives' if not regulated by additional mechanisms. Sources / title words / authors / references particularly associated to a specialty may also feature in publications in other specialties, which as a result could be included as false positives. Some of the publications in a specialty may not feature any source / title words / authors / references particularly associated to that specialty, and could as a result remain un-included as false negatives.

The proposed method aims to accommodate a specialty's internal similarity as well as diversity, and limit false positives as well as false negatives. It does so by requiring coverage by key values for not just one, but at least three of the four data fields. The 'required' coverage by key values for 'at least' three of the four data fields strengthens similarity and limits false positives. The 'allowed' coverage by key values for 'only' three of the four data fields accommodates some diversity and limits false negatives.

A different way to look at this mechanism is as a way to enlarge the specialty approximation beyond those publications covered by key values for all four data fields. Also publications not covered by key values for a particular data field (featuring only values that are not or less strongly or only recently associated to the particular specialty, or associated to the specialty but not present in the seed record) are included if covered by key values for all three other data fields. The mechanism thus includes in a controlled, selective way new or less frequently occurring elements associated to the specialty next to established or more frequently occurring elements. It thereby accommodates the inclusion of publications reflecting dynamics at the specialty's boundaries and emergence of novelties (or the 'adjacent possible', as discussed in a context of creativity and linguistics by Loreto et al., 2016), important for applications where the possibility to explore the boundaries and dynamics of a specialty is an important aspect.

The legitimacy of these (partial) combinations of key values can be understood from the strong interconnectedness of the related publication data fields. Values in all four data fields are the result of conscious, coordinated actions by authors when preparing a publication: in their capacities of writers choosing subjects and words, linkers choosing documents referred to, submitters choosing journals and other sources submitted to, and collaborators choosing other scientists to work and publish with (Borgman and Furner, 2002). This interconnectedness of particular values from different data fields has for instance been observed already in an early indexing context in co-occurrences of title words and cited references (Small, 1973a). Because of this interconnectedness, combining key values for several data fields strengthens the selection towards the specialty aimed for, as an alternative to more stringent criteria for one data field.

Practically, the specialty approximation can be assembled from four subsets, each containing the publications covered by key values for a different combination of three of the four data fields. Characteristics of the specialty approximation can be expected to be similar to but not necessarily identical to those of the seed record, which are possibly more influenced by local context or individual choices. The specialty approximation does not perfectly render a specialty, but approximates it. For certain applications relatively uncomplicated threshold definitions and key value operationalizations such as those used in this paper may suffice (as illustrated in the results section testing the ability to identify peers). For other applications different or more advanced definitions and operationalizations may be more suitable. Realistically however, even in the most advanced set-up, false positives and false negatives with respect to what is aimed for are still possible.

**4. Application and data**

*4.1. Test of the method's focus on the specialty aimed for*

From the different potential areas of application (of which examples are given in the discussion section), one was chosen to illustrate how the method works and at the same time test its focus on the specialty aimed for. The chosen application is the identification of potential reviewers for an individual scientist's dossier, and the test material is a sample of known peers stemming from a real peer review procedure (a fellowship awardee and four reviewers suggested by the awardee). This type of sample (an applicant and suggested reviewers) is particularly suitable for the test:
- The suggested reviewers' publication records can be trusted to (at least partly) belong to the same specialty as that of the applicant and therefore should be connected by an adequate approximation thereof: The applicant and suggested reviewers are recognized to belong to a same specialty by an expert in that specialty (the applicant). To this respect, it should be noted that an applicant might suggest certain reviewers for their expertise in a particular aspect of the application, who do not necessarily have a same level of expertise in other aspects of it.
- The suggested reviewers' publication records are largely disjoint from the publication record of the applicant and therefore not associated in a trivial way to a same specialty or its approximation (as would be the case for strongly overlapping publication records): Reviewers cannot be hindered by conflicts of interest, as may be evident for instance from (at least recent) co-publications. To this respect, it should be noted that the suggested experts might therefore not include those peers whose expertise is situated closest to the applicant's, but peers whose expertise is situated still close enough to be able to assess the quality of the research proposal and the scientific performance level of the applicant.

More in particular, it is tested whether two requirements for the method to be successful are satisfied:
[R1] the specialty approximation constructed for the fellowship awardee should contain publications of known established peers such as the four suggested reviewers, and
[R2] in particular the most prominent authors in the specialty approximation constructed for the fellowship awardee should be scientists who can be confirmed to belong to the same specialty.

The results section subsequently illustrates the method's three phases including findings in response to these requirements. Finally, it is verified how characteristics of the publication records of a group of peers detected via the specialty approximation relate to those of the group of known peers from the peer review procedure.

*4.2. Policy relevance of the illustrated application*

The strong focus on grants and awards for individual scientists in local, national and international research policies makes the identification of peers to an individual scientist (as potential reviewers for the evaluation of such applications) a relevant choice to illustrate the method and its potential. In peer review procedures the available information to start from typically consists of the submitted applications, possibly accompanied by suggested reviewers. Also when such suggestions can be used for (part of) the invited reviews, the evaluation coordination needs to be able to identify complementary or alternative potential reviewers. The coordination should be free of conflicts of interest, and is typically assigned to an administrator or to a scientist active in the related broad domain, but not in an applicant's specialty. In general, also more potential reviewers need to be identified than strictly needed due to declines. Identifying potential reviewers can therefore be a difficult and time-consuming task that would welcome adequate support from bibliometric methods. In response to this need, new services and tools to support the identification of reviewers for grant evaluation have been proposed by several companies and platforms in recent years. In this developing context the proposed method could be a basis for one possible approach.

*4.3. The investigated sample of known peers*

The investigated sample of scientists is situated in the Biology discipline and consists of an awardee of a senior research fellowship (*FA*) and four experts who were suggested by the awardee as potential reviewers for the application (*SR*(1..4)). The senior research fellowship funding scheme of the Vrije Universiteit Brussel offers the opportunity to apply for a 5-year senior research fellowship in succession of a 5-year European Research Council Starting Grant. The latter is a high-level grant awarded to researchers with 2-7 years of experience since completion of PhD. During this external grant, the grantee holds a tenure-track or tenure position at the university. Consequently, around the time of application for succession by a senior research fellowship, applicants can be in a range of 7-12 years of experience since completion of PhD, and at the end of a 5-year tenure-track position or in tenure. The senior research fellowship awardee is therefore in an advanced career phase, which can also be expected of the suggested reviewers.

Publication data from a 10-year period until the year of application (2012) were collected from the online WoS for each scientist in the sample, including all document types. **Table 2** shows the scientists' different professional environments and publication profiles. The suggested reviewers are found to publish to a varying extent in sources and cells that are in common with the fellowship awardee. This may follow from the individual scientists' differing research niches and research levels (Narin, Pinski, and Gee, 1976), contexts (e.g., institutional, national), and personal choices (Sugimoto and Cronin, 2012). Experts (possibly partly) active in a same specialty thus do not necessarily publish a large share of their publications in sources or even cells that are in common. Vice versa, scientists that do publish a large share of their publications in cells and even sources that are in common do not necessarily belong to a same specialty (Rons, 2016).

The reviewers suggested to evaluate the proposal for the applicant's own research are expected to be experts concerning at least part of the proposed research in the applicant's specialty. In the actual evaluation process, reviews from suggested reviewers form only a part of the invited reviews. In this paper's analysis only the suggested reviewers are included. Their suggestion by the applicant bears a strong guarantee that they are active in the applicant's specialty, which is an important aspect for their usability for the described test of the method. As an extra confirmation that the fellowship awardee and the suggested reviewers are active in the same specialty, a bibliometric or other connection between them at the specialty level is added to **Table 2**.

**Table 2**

Key figures for the publication records and seed records of the fellowship awardee and suggested reviewers.

|  | FA | SR(1) | SR(2) | SR(3) | SR(4) |
|---|---|---|---|---|---|
| $I$ | University | University | University | Museum | University[a] |
| $P$ | 30 | 45 | 38 | 10 | 226 |
| $P^s$; $P^s/P$ | 2-13; 7-43% | 12; 27% | 9; 24% | 5; 50% | 7; 3% |
| $P^c$; $P^c/P$ | 17-20; 57-67% | 15; 33% | 22; 58% | 7; 70% | 9; 4% |
| $Ar_P$ | 27 | 39 | 36 | 2 | 174 |
| $Re_P$ | 0 | 1 | 1 | 4 | 20 |
| $MA_P$ | 1 | 2 | 0 | 0 | 22 |
| $EM_P$ | 0 | 2 | 1 | 4 | 3 |
| $Le_P$ | 2 | 1 | 0 | 0 | 5 |
| $PP_P$ | 0 | 0 | 0 | 0 | 1 |
| $Co_P$ | 0 | 0 | 0 | 0 | 1 |
| $S$ | 262 | 483 | 357 | 182 | 1861 |
| $Ar_S$ | 209 | 381 | 298 | 149 | 1587 |
| $Re_S$ | 39 | 73 | 45 | 21 | 220 |
| $EM_S$ | 6 | 21 | 10 | 8 | 13 |
| $Le_S$ | 4 | 2 | 2 | 3 | 17 |
| $MA_S$ | 1 | 2 | 0 | 0 | 22 |
| $SR_S$ | 2 | 2 | 0 | 0 | 0 |
| $NI_S$ | 0 | 1 | 1 | 1 | 0 |
| $PP_S$ | 0 | 0 | 1 | 0 | 1 |
| $BR_S$ | 0 | 1 | 0 | 0 | 0 |
| $Rp_S$ | 1 | 0 | 0 | 0 | 0 |
| $Co_S$ | 0 | 0 | 0 | 0 | 1 |
| Connection[b] |  | [I] | [II] | [III] | [IV] |

FA: Fellowship awardee (Discipline: Biology).
SR(*i*): Suggested reviewer *i*.
*I*: Type of institutional affiliation.
*P*: Number of publications in Thomson Reuters Web of Science (WoS) Core Collection, 2003-2012.
$P^s$, $P^c$: Numbers of publications in *P* respectively in sources and cells in common between FA and SR(*i*)
*S*: Number of publications in the seed record, in Web of Science (WoS) Core Collection, 2003-2012.
$Ar_j$, $Re_j$, $EM_j$, $Le_j$, $MA_j$, $SR_j$, $NI_j$, $PP_j$, $BR_j$, $Rp_j$, $Co_j$: Number of publications in *j* (*P* or *S*) of respectively document type Article (incl. Article and Proceedings Paper), Review (incl. Review and Book Chapter), Editorial Material, Letter, Meeting Abstract, Software Review, News Item, Proceedings Paper, Book Review, Reprint, Correction.
[a]: Department in the biomedical sciences domain.
[b]: Scientific connection at the specialty level to the fellowship awardee.
[I]: Specialized co-publications in WoS (until 6 years before the year of application).
[II]: Cited in same specialized publications.
[III]: References in same specialized Wikipedia page.
[IV]: Specialized co-publications in WoS (after the year of application).
*Data sourced from Thomson Reuters Web of Knowledge (formerly referred to as ISI Web of Science).*
*Web of Science accessed online 29-30.09.2015, and 31.03.2016*

**5. Results**

*5.1. Results in Phase 1: Seed records*

The publication records of the fellowship awardee and of the suggested reviewers were enlarged to seed records for further analysis by adding the publications referred to, as described in the method sub-section for Phase 1. **Table 2** shows that the seed records are larger and more uniform in terms of distributions over document types compared to the publication records of the individual scientists.

*5.2. Results in Phase 2: Key values*

Key values were determined as described in the method sub-section for Phase 2 for the fellowship awardee and for the four suggested reviewers. A same relatively large coverage threshold of 80% was used to determine each set of key values, aiming for a substantial coverage of the specialty by the specialty approximation within which the presence of known peers is to be verified. The same threshold value was used by Rons (2016) aiming for a substantial coverage of two closely related specialties by the respective specialty approximations for which the degree of separation was to be verified. **Figure 2** shows how for each of the five scientists (not individually identified in this figure) and for each of the four data fields relatively small fractions of key values among unique occurring values (horizontal axis) were found to cover at least 80% of the seed records' publications (vertical axis).

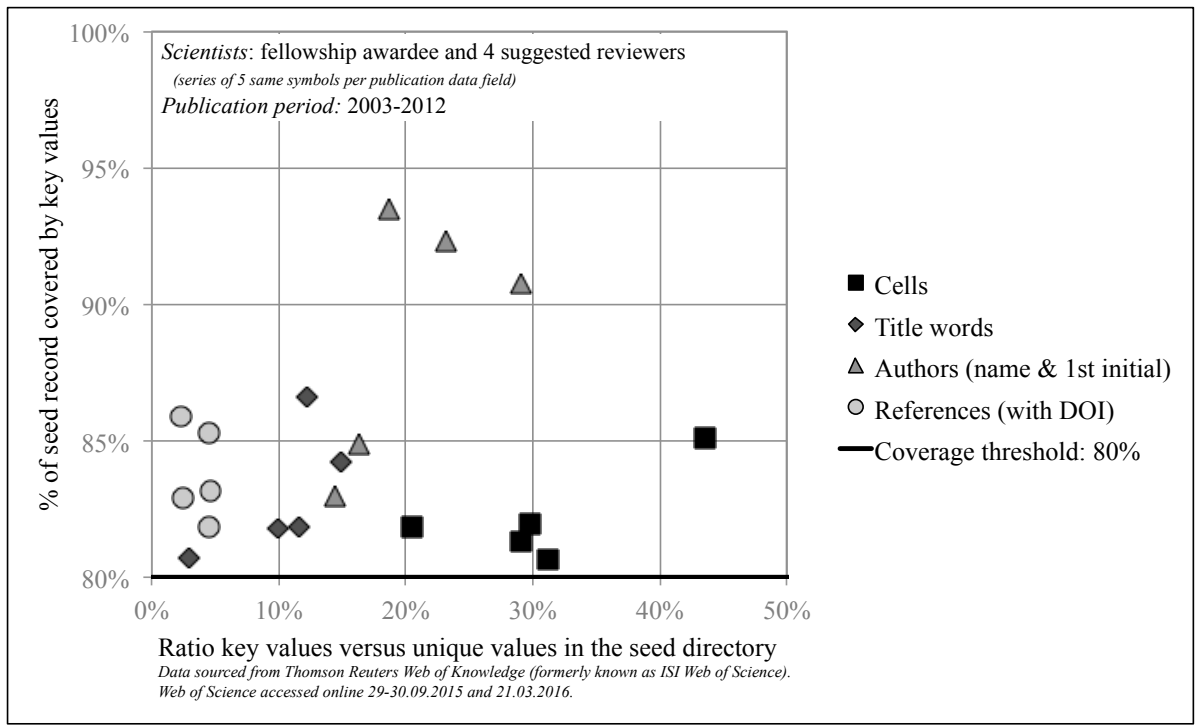

**Fig. 2.** Ratios of key values versus unique values, determined for the fellowship awardee and suggested reviewers.

With each set of key values covering at least 80% of publications in the seed record, a seed record largely consists of publications covered by key values for at least three of the four data fields, satisfying the requirements to be included in the specialty approximation. These publications would constitute 82% of the seed record in the imaginary case where the values in the four data fields would be independent (where seed record publications would be included in the specialty approximation with a probability of $4 \times (80\%)^3 \times (1-80\%) + (80\%)^4 = 82\%$). In reality the four publication data fields are not independent (see section 3.3. for a discussion on the interconnectedness) and the percentage of seed record publications included in the specialty approximation can be expected to be larger than 82%. The remaining seed record publications are covered by key values for only two, one or none of the four data fields. This is illustrated schematically in **Figure 3** for publications in the fellowship awardee's seed record, ordered on the vertical axis in five shares (separated by dashed lines) according to coverage by key values (covered: light grey; not covered: dark grey), starting at the bottom with the largest share of publications covered by key values for all four data fields (65%), followed by the share of publications covered by key values for three of the four data fields (24%), for two of the four data fields (8%), for one of the four data fields (2%) and for none of the four data fields (0.4%). The first two shares joined together contain the large majority of the seed record's publications that is covered by key values for at least three of the four data fields (89%, larger than 82% as expected).

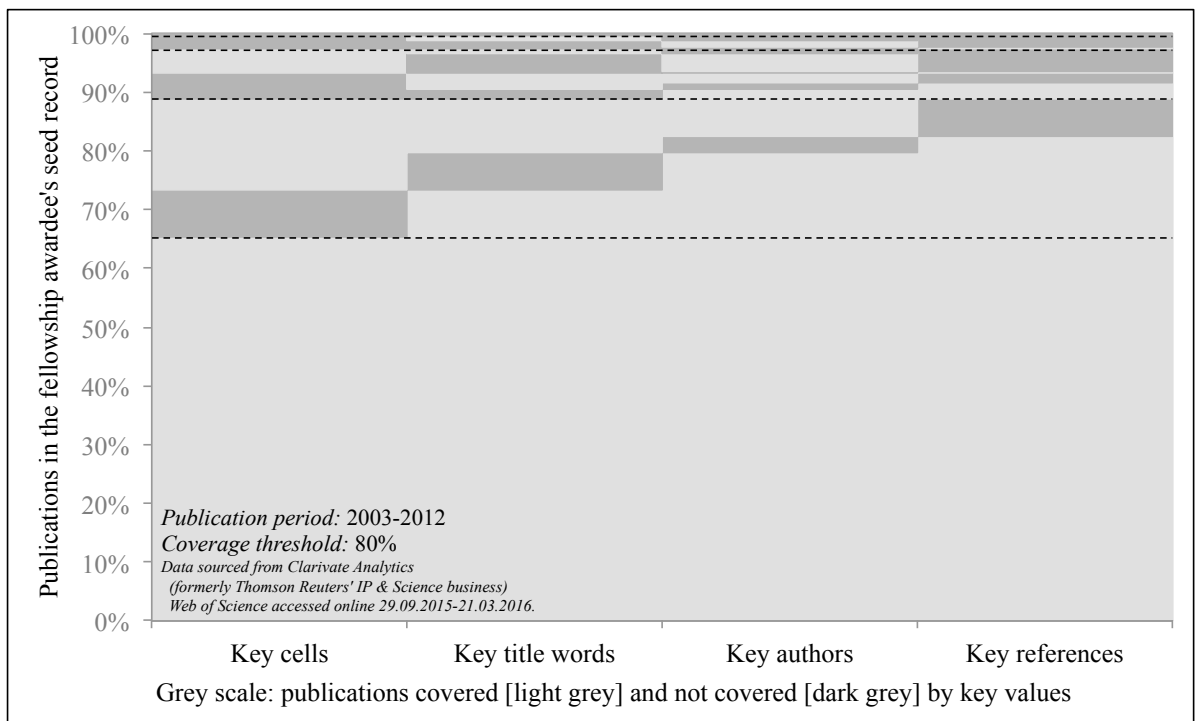

**Fig. 3.** Schematic presentation of the coverage of the fellowship awardee's seed record by key values determined from it.

When the method is applied to identify potential reviewers, the set of key authors offers a first opportunity. From the four suggested reviewers, two belong to the set of key authors determined for the fellowship awardee. It would have been possible therefore to detect these two already in this phase, before constructing the actual specialty approximation.

To verify whether publications of the four suggested reviewers are present in the specialty approximation for the fellowship awardee [R1], the specialty approximation itself does not need to be constructed. It can simply be verified whether parts of the publication records of the suggested reviewers are covered by key values for at least three of the four data fields. This is the case for all four of the suggested reviewers, who were found to be present in the fellowship awardee's specialty approximation with strongly varying shares of their publications: few of the publications of *SR*(2) and *SR*(4) and a majority of the publications of *SR*(1) and *SR*(3) (percentages on the horizontal axis of **Figure 6**).

This variation can be understood from the method's aim, which is not to exhaustively include all of the publications in the specialty by any scientist belonging to it (such as a suggested reviewer), but to generate an approximation of a specialty that is sufficiently focused for usage in a particular research policy context, in the examined case the identification of potential reviewers. Equally valid reviewers can be more or less prominently present in a constructed specialty approximation, and their publication records more or less well covered by it, depending on personal environments and choices. An affiliation to a department with different publication characteristics and topical focus compared to the specialty (for example a statistician working in another science department than mathematics) may result in a publication record shaped accordingly (possibly impacting for instance publication volume and venues), with possibly smaller fractions of it covered by a specialty approximation constructed for a seed record from a more common departmental environment for the specialty. This is the case for suggested reviewer *SR*(4), who is situated in a department in the biomedical sciences domain and whose publications study a large range of biological classes (one of which is studied by the fellowship awardee) from a medical perspective. Membership of a hyperspecialized research community on the other hand, with a specific group of authors and body of literature, may also result in a publication record of which a smaller fraction is covered by a specialty approximation constructed for a seed record with a broader representation of the specialty. This is the case for suggested reviewer *SR*(2), whose publications focus on a particular order in the biological class studied by the fellowship awardee. Such wider or narrower scopes do not prevent scientists to suggest potential reviewers able to evaluate a dossier as a whole or particular aspects of it, based on their knowledge of the research community and on their understanding of the particularities of an individual cv or dossier. This observed variation in bibliometric profiles among peers illustrates the complexity of the specialty concept, stretching out towards other specialties and hyperspecializing within.

*5.3. Results in Phase 3: Specialty approximation*

The full specialty approximation (8172 publications) was constructed for the fellowship awardee as described in the method sub-section for Phase 3, for the last three years of the observed publication period until the year of application (2010-2012). This time frame results in a focus on active authors around the time of application, among whom reviewers would be sought.

**Figure 4** schematically shows the specialty approximation's composition, with publications ordered on the vertical axis in two shares (separated by a dashed line) according to coverage by key values (covered: light grey; not covered: dark grey). Corresponding to its definition, the specialty approximation contains publications covered by key values for each of the four data fields (a minor share at the bottom) and publications covered by key values for the four possible combinations of three of the four data fields (the large majority). The latter for a large part consists of publications not covered by key authors, which can be understood from the origin of the seed record. The more exhaustively the prominent cells, title words, authors or references associated to the specialty already figure in a seed record, the less the specialty approximation will contain publications that are not covered by the respective key values determined from the seed record. The results show that, even when an enhanced seed record is constructed starting from the publication record of an individual scientist, it may still be limited in its inclusion of prominent authors, resulting in a relatively large share of publications in the specialty approximation not (co-)authored by any of the key authors determined from the seed record.

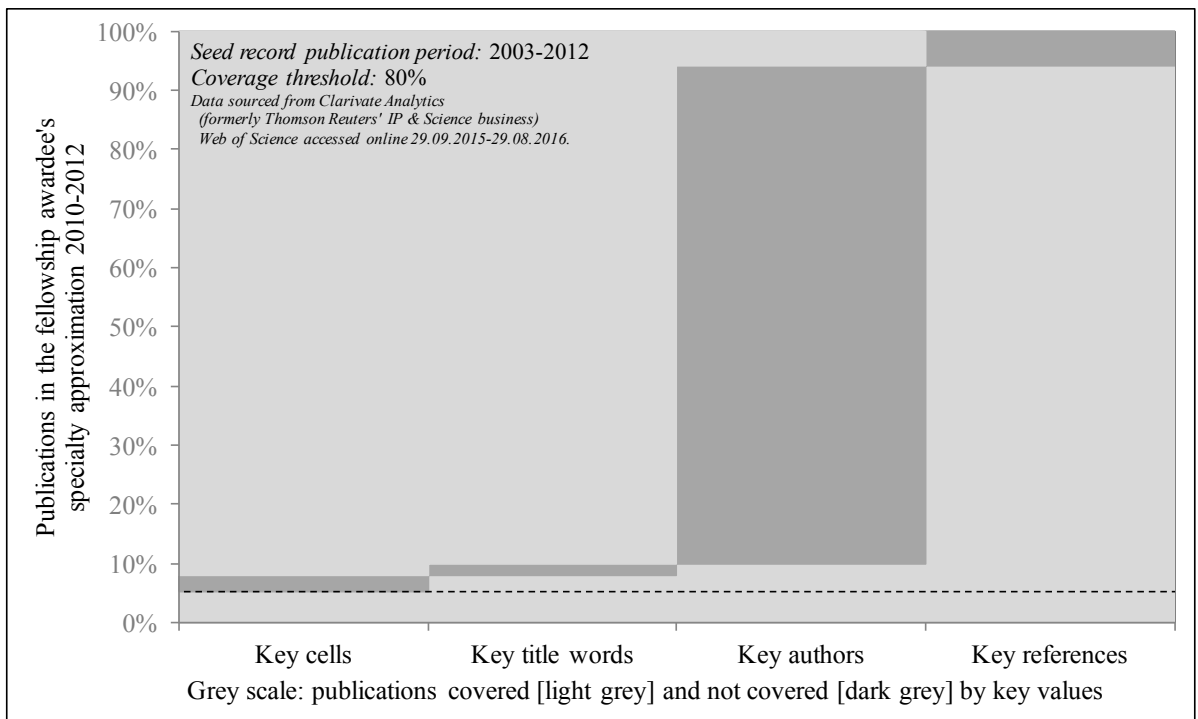

**Fig. 4.** Schematic presentation of the coverage of the fellowship awardee's specialty approximation by key values used in its construction.

If the constructed specialty approximation effectively approximates a specialty, then a relatively large share of its publications should be covered by a relatively small share of its authors, as well as references, title words and cells. To verify this, new sets of key values ('key values*') were determined starting from the constructed specialty approximation as the new seed record ('seed record*'). Similar proportions were indeed obtained for the coverage of the constructed specialty approximation by key values* (**Figure 5**) as for the coverage of the initial seed record by the initial key values (**Figure 3**). The ratios of key values* versus unique values were found to be similar or smaller compared to the same ratios calculated for the initial key values determined for the five scientists in the test sample (shown in **Figure 2**, horizontal axis): 21% for authors, 8% for cells, 0.3% for title words and 0.01% for references. A more detailed analysis of the nature of publications in the constructed specialty approximation is out of scope of this paper, and not required for the demonstrations made. It may however be essential when thresholds and criteria definitions need to be tuned to suit particular applications.

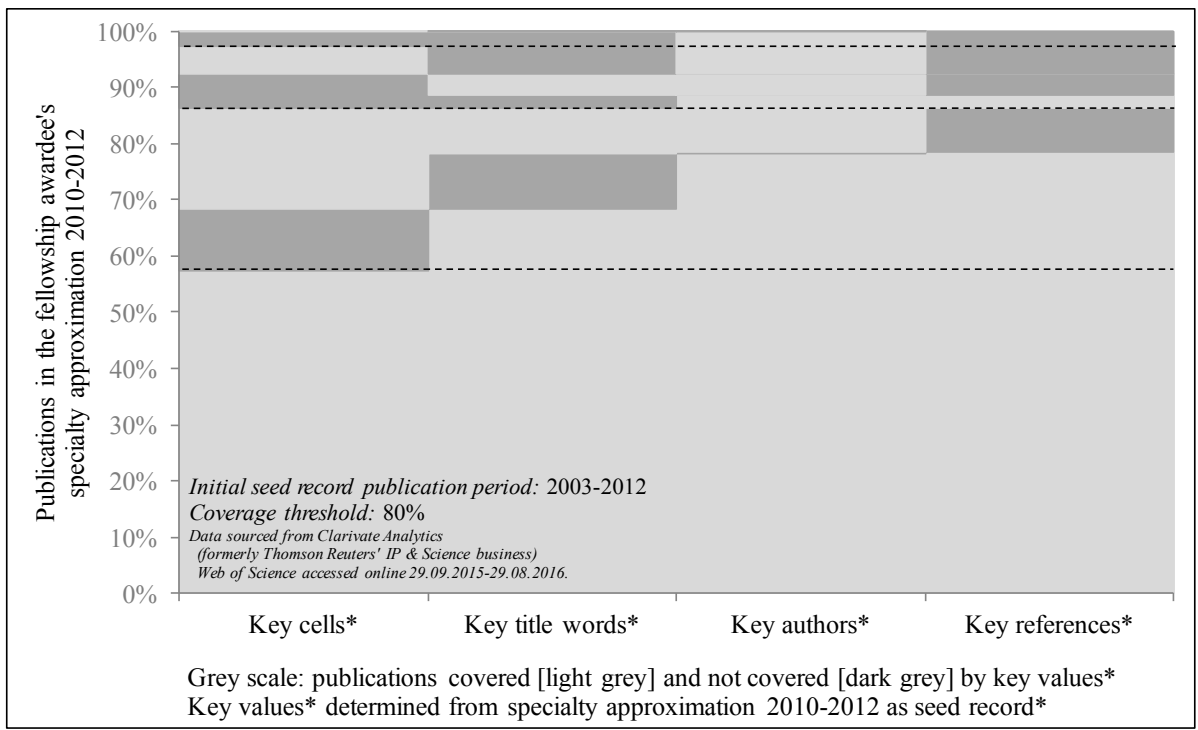

**Fig. 5.** Schematic presentation of the coverage of the fellowship awardee's specialty approximation by key values* determined from it.

*5.4. Potential reviewers identified via the specialty approximation method*

Potential reviewers can be identified in two phases of the proposed method: already among the key authors or in the constructed specialty approximation. For the illustrations in this paper five potential reviewers for the fellowship awardee were selected among the most prominent authors in both phases: three from key authors and three from the constructed specialty approximation, one figuring in both selections. Both selections of prominent authors are based on absolute numbers of publications (as one of the possible criteria for such a selection) and exclude the fellowship awardee and authors with co-publications (who would be disqualified in a search for reviewers).

Co-authors can be expected to occur more frequently in the set of key authors, in general situated closer to a scientist's personal environment than the set of authors present in the entire specialty approximation. This expectation is confirmed by results for the fellowship awardee, where 6 co-authors are present among the 11 most frequently occurring key authors and only 1 co-author is present among the 11 most frequently occurring authors in the specialty approximation (the most frequently occurring authors excluding the fellowship awardee in both cases). For this comparison, co-authorships were determined via the Web of Science in the publication period 2003-2012. In any practical application of the method to help identify potential reviewers, suggested experts would need to be tested for conflicts of interest (resulting for instance from co-publications) as is prescribed or customary for the particular evaluation, independent of how suggested reviewers are obtained.

If the constructed specialty approximation is sufficiently focused on the specialty aimed for, at least its most prominent authors should be active in that specialty [R2]. For the co-authors among them, the co-publications signal activity in the same specialty together with the fellowship awardee. To verify whether the five selected prominent authors without such co-authorships can be confirmed to be active in the same specialty as the fellowship awardee, non-bibliometric scientific connections between them at the specialty level were sought and found (**Table 3**).

**Table 3**

Publicly available non-bibliometric scientific connections at the specialty level between the fellowship awardee and prominent authors selected from key authors and from the constructed specialty approximation.

| Author | A(1) | A(2) | A(3) = B(1) | B(2) | B(3) |
|---|---|---|---|---|---|
| Connection | [I] | [II] | [II], [III], [IV] | [III] | [V] |

[I]: Contributors at a same specialized congress programme, with A(1) as plenary speaker
[II]: References in same specialized Wikipedia pages
[III]: Contributing specialized scientists on a same list maintained by an international union
[IV]: Lecturers in a same specialized Master programme
[V]: Reviewers acknowledged on a same list by a specialized journal
A(i) and B(i): Top 3 most frequently occurring authors identified among key authors (A(i); A(1) and A(2) ex aequo) and from the constructed specialty approximation (B(i)) by unique name & 1st initial, avoiding homonyms by excluding frequently occurring names, disregarding the awardee and authors connected to the awardee by co-publications in the 5 years until the year of application (2012).

*5.5. Comparison with reviewers suggested by the fellowship awardee*

In this subsection it is shown that the group of potential reviewers identified via the method is not less well connected bibliometrically to the fellowship awardee than the group of suggested reviewers. For members of both groups if was investigated to which extent (1) their publications are present in the fellowship awardee's specialty approximation and (2) their key values cover the fellowship awardee's specialty approximation. **Figure 6** and **Figure 7** show how both groups share a similar variability and include members with high and low scores.

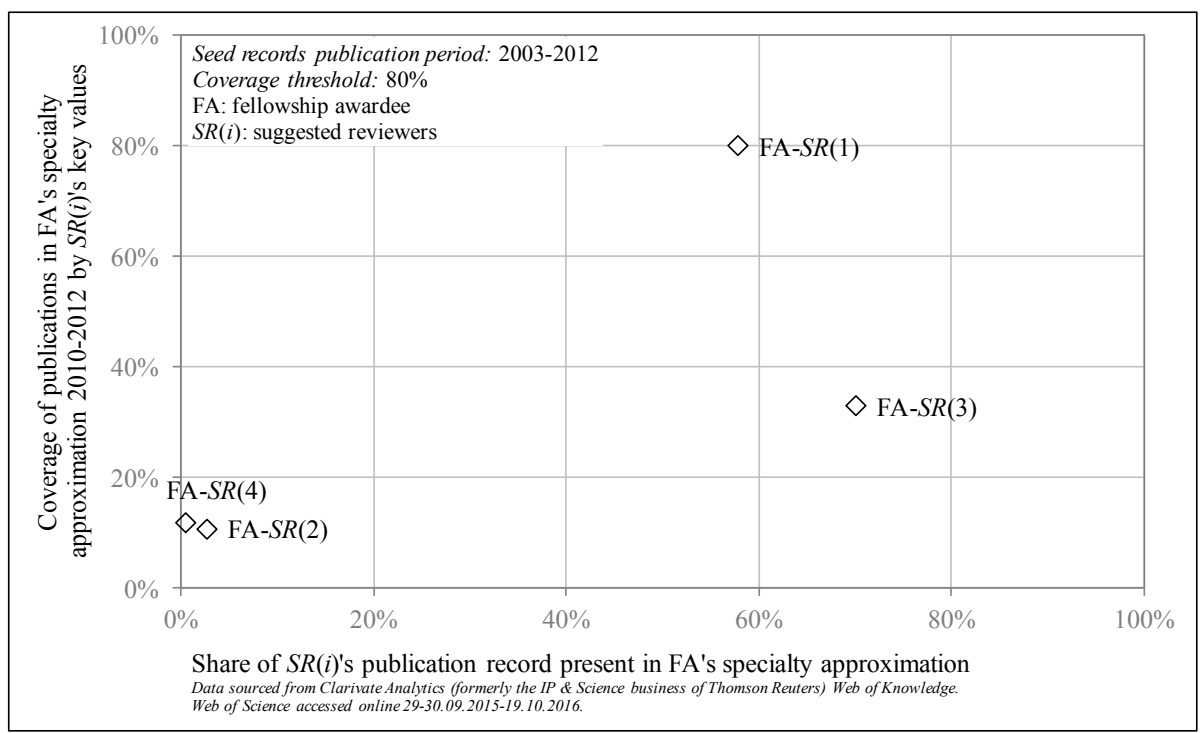

**Fig. 6.** Mutual coverage levels: fellowship awardee versus suggested reviewers.

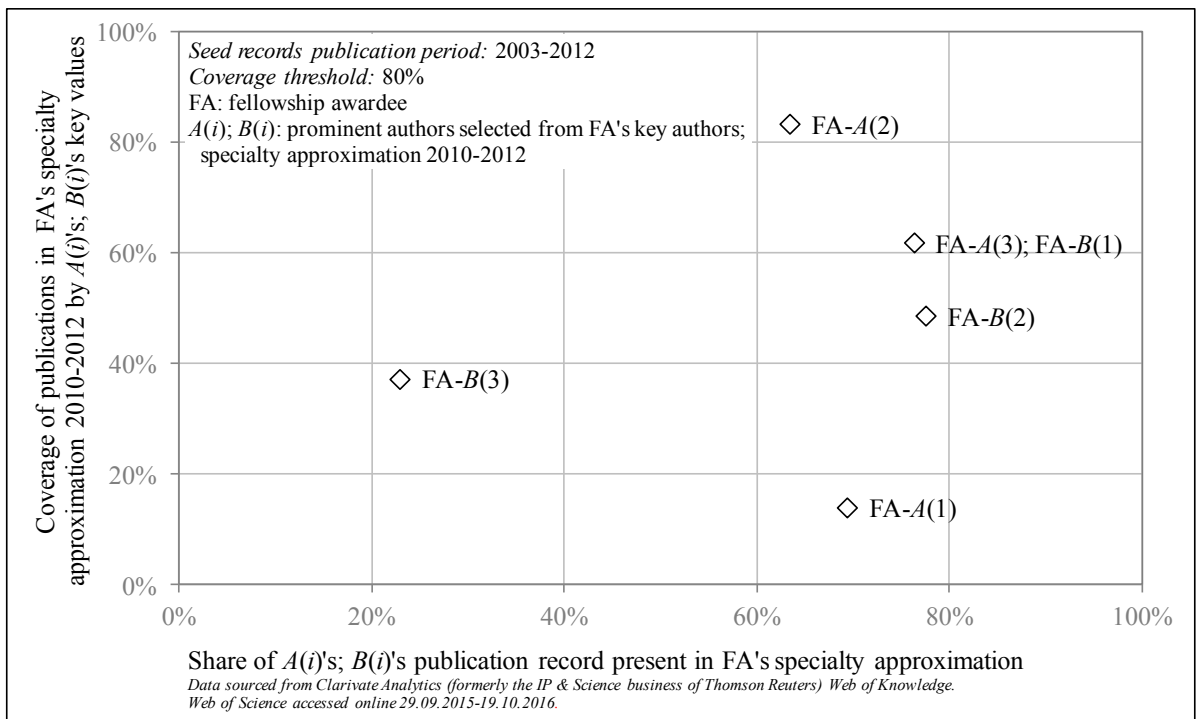

**Fig. 7.** Mutual coverage levels: fellowship awardee versus prominent authors selected via the specialty approximation method.

## 6. Discussion

*6.1. Test of the method's focus on the specialty aimed for*

The application shown in this paper at the same time illustrates the method's results and tests its focus on the specialty aimed for in terms of its ability to connect and identify peers. The results obtained for the test sample indicate that potential reviewers for a scientist's work can be identified via the constructed specialty approximation and via key authors used in its construction (the latter likely to include proportionally more co-authors). The set of prominent authors selected from the method consisted of experts who could be confirmed to be active in the scientist's specialty, and its bibliometric characteristics do not deviate from those observed for the set of reviewers suggested by the scientist. Both sets of experts encompass a similar variability in mutual coverage levels. The method is thus able to connect peers and provide a basis from which to identify and recommend peers as one possible application, at least in the investigated case. Additional studies are needed to further confirm this ability for a wide spectrum of specialties and for interdisciplinary research, possibly involving adapted method designs.

*6.2. Potential for more advanced method designs*

This paper describes simple operationalizations, adequate for this 'proof of concept' paper introducing the method and for the illustrated application. The obtained results show that these simple operationalizations can already yield sufficiently adequate results for certain applications, such as the identification of peers. Designs can be refined in each phase of the method to better fit particularities of a studied specialty or of an application. More advanced designs or techniques could be used to improve the resulting accuracy of delineation, or the extent to which the specialty aimed for is represented. Certain applications may require such higher precision and completeness levels (e.g., creating reference sets for bibliometric indicators) and possibly also a restriction of results to for example a particular document type or language (Sirtes, 2012; van Raan, van Leeuwen, and Visser, 2011).

Examples of more advanced designs for the considered values for the four data fields are given below:
- Source field: In domains where sets of specialty-related sources are available that are validated by for example scientific associations, this local source-structure could be used instead of the more globally available subject category structure.
- Title field: Stop words could be excluded in a more advanced way by excluding domain specific stop words with no discriminant value within the specific domain (Makrehchi and Kamel, 2008), determined from abstracts or full text. Also key phrases could be identified from titles and merged in a common list with key title words, as proposed by Milojevic et al. (2011). Where abstracts or full text are generally available for an area studied, more advanced criteria derived from Natural Language Processing techniques could substitute the criterion based on title words (e.g., resembling techniques to derive sets of terms characterizing fine-grained clusters of publications as used by Waltman and Van Eck, 2012).
- Authors field: Author disambiguation is a challenge in particular where an important part of the contributions to a specialty stems from geographic regions with high homonym frequency (e.g., China, Korea). This challenge also holds at the disciplinary level, as investigated for Chemistry, Physics, Medicine, and Economics and Business by Harzing (2015). Even if the problem would be less pronounced at the specialty level, results may still benefit from more advanced disambiguation methods (Schulz, 2016).
- References field: In particular in analyses of specialties for which an important part of the common knowledge base may not be covered by the publication and citation database used, references to publications not indexed in the database (Butler and Visser, 2006) could be taken into account.

Also for the selection and subsequent processing of key values this paper describes a simple approach where all data fields and all key values for a particular data field are treated equally. It could be explored whether variants with more differentiation can be tuned to obtain better results for particular applications.

*6.3. Potential areas of application*

The proposed method can be characterized as a 'bottom-up' approach, constructing a specialty approximation starting from a particular specialized seed record. This constitutes an important difference in view of potential applications compared to 'top-down' techniques that cluster large publication sets into fields or topics (e.g based on words via Latent Dirichlet Allocation: Blei, Ng, and Jordan, 2003; or based on citation relations: Waltman and van Eck, 2012 and Boyack and Klavans, 2014, the latter also using textual similarity for visualization in a map). The bottom-up aspect strongly determines the method's possibilities and limitations.

Tests performed until now indicate that, at least for the examples investigated, the method is able to generate specialty approximations that connect peers (this paper), and distinguish between closely related specialties (Rons, 2016). These abilities and the required seed record as a starting point situate the method's potential utility in particular in the ad hoc aggregation of specialty approximations, and in the exploration of contributors therein.

**Table 4** gives some examples of potential bibliometric areas of application seeking to produce or investigate a constructed specialty approximation as such, contributors to it, or coverage by it. The way in which criteria are combined, requiring most but not all to be satisfied, may also be applicable in other contexts where coherent groups of items are represented by interrelated values in several dimensions.

**Table 4**

Examples of potential bibliometric areas of application for the proposed specialty approximation.

| Type of result produced or investigated | Potential area of application |
|---|---|
| A constructed specialty approximation | - Research profiling / literature review / description of prior art<br>- Recommendation of literature<br>- Revealing differences between specialties<br>- Situation in a domain of reference |
| Contributors to a constructed specialty approximation | - Recommendation of experts, reviewers, benchmarks<br>- Recommendation of collaborations, mobility, internationalization<br>- Identification of important contributors<br>- Monitoring of contributors |
| Coverage by a constructed specialty approximation | - Confirmation of relevant expertise<br>- Indication of expertise overlap<br>- Estimation of programme coverage by proposals |

Some potential bibliometric areas of application have been contributed to for some time by other techniques (e.g., recommendation of literature, situation in a domain of reference, recommendation of collaborations) and it could be investigated whether and how the proposed specialty approximation can offer a sufficiently adequate alternative domain delineation.

For certain potential applications such as the recommendation of collaborations, a focus in literature has been on global policy perspectives, pointing to best options for collaborations at research topic level from a range of disciplines and institutions (Boyack, 2009) or for regional collaboration from a range of cities (Guns and Rousseau, 2013). From the perspective of scientists in a particular specialty, the added value of bibliometrically generated recommendations of collaborations may at first sight seem small, as its leading scientists can be expected to be able to identify suitable collaborations based on their knowledge of their own research community. Still, bibliometric tools may be helpful in certain circumstances such as extending collaborations towards less familiar areas, or in interdisciplinary research.

Some potential applications such as the level of panel expertise (Rahman et al., 2015) and evaluator distance from an evaluated dossier (Boudreau et al., 2016) have gained attention more recently. These areas can lead to bibliometric instruments that support peer review procedures in a new way: contributing information concerning (potential) reviewers instead of the reviewed. Further studies in this direction could result in improvements in the domains of selection of reviewers, interpretation of peer review results, and understanding of underlying mechanisms driving peer judgement in relation to intellectual distance.

*6.4. Issues in comparative contexts*

Certain potential applications may seek to compare multiple subjects in possibly multiple specialties. In particular in such contexts it should be stressed that relative positions in different specialty approximations (or even in a same one) need to be interpreted with caution. Also in specialty approximations that are more closely fit to a specialty than broader approximations (for example subject category based), certain differences between particular subjects and specialties that complicate a comparative interpretation persist (Yuret, 2015). In particular when studying publication records of individual scientists, differences in career phase, context (national, institutional), and individual choices in publication behaviour may influence bibliometric results.

Some differences between investigated subjects can be made apparent relatively easily in bibliometric analyses, for example differences in coverage by the database used as estimated from a study of references. Other differences can be easily discerned and pointed out by peers, but are less straightforward to reveal in a bibliometric analysis, for example differences in balance of research levels (basic/applied/empirical/clinical) and in research focus. These may however be associated to different typical bibliometric characteristics, and influence values of indicators that are based on these characteristics. Concerning balance, specialties may encompass contributions from different levels of research (Narin, Pinski, and Gee, 1976), characterized by (partly) different words, publication venues, (numbers of) authors, institutions, and literature referred to (Lewison and Paraje, 2004; Tijssen, 2010; Boyack et al., 2014), and by different citation impact levels (van Eck et al., 2013; Johnston, Piatti and Torgler, 2013). Concerning focus, specialties may encompass different subareas, for example dedicated to a specific natural species or medical treatment. Global contextual factors may significantly guide explorations and investments towards specific focal areas, for example a privileged strategic priority or societal challenge (Bos et al., 2014), a hype of expectations (van Lente, Spitters, and Peine, 2013), or a difficult to challenge established tradition or leadership (Azoulay, Fons-Rosen, and Graff Zivin, 2015). The resulting growth presents an advantage compared to more static areas in terms of citations (Levitt and Thelwall, 2016).

*6.5. Target scale and scalability*

The method was designed and tested for application at the level of research specialties such as those of individual scientists or highly specialized research teams or programmes. Depending on the size of the research community associated to a specialty, also the amount of information to be processed will vary (compare for example the different orders of magnitude of the two Particle Physics subdomains investigated by Rons, 2016). Technically, given the required calculation capacity, the same procedure could be applied starting from more broadly defined seed records spread out over a domain overarching a few or many specialties. The resulting domain approximation however may venture farther beyond the specialties aimed to include (via cross-specialty combinations of key values), and would only be well-balanced for domains not dominated by certain subdomains. Dominance by subdomains is not unlikely. Even between closely related specialties, typical publication characteristics can strongly vary. Precisely this variation is an important incentive to develop bibliometric techniques to sufficiently closely delineate a specialty.

**7. Conclusion**

A method is proposed to bibliometrically approximate the specialty of a given highly specialized publication record based on concepts defining scientific disciplines and scholarly communication, and building on observed statistical regularities in publication data. The method partially combines key values for four bibliometric data fields to create a specialty approximation from information available in global publication and citation databases (source, title, authors and references). Results shown to illustrate the method demonstrate how simple operationalizations can generate sufficiently adequate results for at least certain applications and also successfully test the method's focus on the specialty aimed for in terms of its ability to connect and identify peers. If confirmed for a wider range of specialties, this ability could contribute to new types of bibliometric support, for example providing information concerning evaluators instead of the evaluated in evaluation procedures. Some possible alternative operationalizations and potential issues requiring attention were pointed at in the description and discussion of the method. Besides investigations for a wider range of research specialties and more advanced operationalizations, also applications to different kinds of specialized publication records and exploration of the potential for diverse applications in research and research policy context are among the possible tracks for further investigations.

**Acknowledgements**

The author would like to thank the anonymous reviewers and the editor for their constructive comments and suggestions.
**References**

Azoulay, P., Fons-Rosen, C., & Graff Zivin, J.S. (2015). Does Science Advance One Funeral at a Time? *National Bureau of Economic Research (NBER) Working Paper* No. 21788.

Berkenkotter, C., & Huckin, T.N. (1995). *Genre Knowledge in Disciplinary Communication: Cognition/Culture/Power*, Hillsdale, NJ: Lawrence Erlbaum.

Blei, D.M., Ng, A.Y., & Jordan, M.I. (2003). Latent Dirichlet Allocation. *Journal of Machine Learning Research*, 3(4-5), 993-1022.

Borgman, C.L. (1989). Bibliometrics and scholarly communication: Editor's introduction. *Communication Research*, 16(5), 583-599.

Borgman, C.L., & Furner, J. (2002). Scholarly Communication and Bibliometrics. *Annual Review of Information Science and Technology*, 36, 3-72.

Bos, C., Walhout, B., Peine, A., & van Lente, H. (2014). Steering with big words: articulating ideographs in research programs. Journal of Responsible Innovation, 1(2), 151-170.

Boudreau, K.J., Guinan, E.C., Lakhani, K.R., & Riedl, C. (2016). Looking Across and Looking Beyond the Knowledge Frontier: Intellectual Distance, Novelty, and Resource Allocation in Science. *Management Science*, published online in Articles in Advance, January 8, 2016, pp. 1-19. http://dx.doi.org/10.1287/mnsc.2015.2285.

Bourke, P., & Butler, L. (1998). Institutions and the map of science: matching university departments and fields of research. *Research Policy*, 26(6), 711-718.

Boyack, K.W. (2009). Using detailed maps of science to identify potential collaborations. *Scientometrics*, 79(1), 27-44.

Boyack, K.W., & Klavans, R. (2010). Co-citation Analysis, Bibliographic Coupling, and Direct Citation: Which Citation Approach Represents the Research Front Most Accurately? *Journal of the American Society for Information Science and Technology*, 61(12), 2389-2404.

Boyack, K.W., & Klavans, R. (2014). Creation of a Highly Detailed, Dynamic, Global Model and Map of Science. *Journal of the Association for Information Science and Technology*, 65(4), 670-685.

Boyack, K.W., Patek, M., Ungar, L.H., Yoon, P., & Klavans, R. (2014). Classification of individual articles from all of science by research level. *Journal of Informetrics*, 8(1), 1-12.

Braam, R.R., Moed, H.F., & van Raan, A.F.J. (1991). Mapping of Science by Combined Co-Citation and Word Analysis. I. Structural Aspects. *Journal of the American Society for Information Science*, 42(4), 233-251.

Bradford, S.C. (1934). Sources of Information on Specific Subjects. *Engineering: An Illustrated Weekly Journal*, 137(3550), 85-86.

Butler, L., & Visser, M.S. (2006). Extending citation analysis to non-source items. *Scientometrics*, 66(2), 327-343.

Buxton, A. B., & Meadows, A. J. (1977). The Variation in the Information Content of Titles of Research Papers with Time and Discipline. *Journal of Documentation*, 33(1), 46-52.

Callon, M., Courtial, J. P., Turner, W. A., & Bauin, S. (1983). From translations to problematic networks: An introduction to co-word analysis. *Social Science Information*, 22(2), 191-235.

Chen, C., Ibekwe-SanJuan, F., & Hou, J. (2010). The Structure and Dynamics of Cocitation Clusters: A Multiple-Perspective Cocitation Analysis. *Journal of the American Society for Information Science and Technology*, 61(7), 1386-1409.

Chikhi, N.F., Rothenburger, B., & Aussenac-Gilles, N. (2008). Combining Link and Content Information for Scientific Topics Discovery. In: *Proceedings of 20th IEEE International Conference on Tools with Artificial Intelligence, ICTAI, 2008*, pp. 211-214.
N. Rons / Bibliometric Approximation of a Scientific Specialty by Combining Key Sources, Title Words, Authors and References   - 27 / 31 -